\documentclass[twocolumn, tighten]{aastex63}

\usepackage{graphics,graphicx}
\usepackage{epsfig}
\usepackage{color}
\usepackage{xcolor,soul}

\newcommand{\ltsimeq}{\raisebox{-0.6ex}{$\,\stackrel
        {\raisebox{-.2ex}{$\textstyle <$}}{\sim}\,$}}
\newcommand{\gtsimeq}{\raisebox{-0.6ex}{$\,\stackrel
        {\raisebox{-.2ex}{$\textstyle >$}}{\sim}\,$}}

\newcommand{\spit}{\mbox{\it Spitzer}}
\newcommand{\sof}{\mbox{\rm SOFIA}}

\shorttitle{V838~Mon in the IR}
\shortauthors{Woodward et al.}

\DeclareGraphicsExtensions{.png,.pdf,.eps,.ps}
\begin{document}

\title{The Infrared Evolution of Dust in V838 Monocerotis}

\author[0000-0001-6567-627X]{C. E. Woodward}
\affiliation{Minnesota Institute for Astrophysics, University of Minnesota,\\
116 Church Street SE, Minneapolis, MN 55455, USA}

\author[0000-0002-3142-8953]{A. Evans}
\affiliation{Astrophysics Group, Keele University, Keele, Staffordshire,
ST5 5BG, UK}

\author[0000-0002-9670-4824]{D. P. K. Banerjee}
\affiliation{Physical Research Laboratory, Navrangpura, Ahmedabad, Gujarat 380009, India}

\author[0000-0003-2196-9091]{T. Liimets}
\affiliation{Astronomick\'y  \'ustav, Akademie v\v{e}d \v{C}esk\'e republiky, v.v.i., 
Fri\v{c}ova 298, 251\,65 Ond\v{r}ejov, Czech Republic}
\affiliation{Tartu Observatory, University of Tartu, Observatooriumi 1, 61602 T\~oravere, Estonia}

\author[0000-0001-6316-9880]{A. A. Djupvik}
\affil{Nordic Optical Telescope, Rambla Jos\'e Ana Fern\'andez P\'erez 7, ES-38711 Bre\~na Baja, Spain}

\author[0000-0002-1359-6312]{S. Starrfield}
\affiliation{School of Earth and Space Exploration, Arizona State University, 
Box 871404, Tempe, AZ 85287-1404, USA}

\author[0000-0002-0141-7436]{G. C. Clayton}
\affiliation{Louisiana State University, Baton Rouge, Department of Physics \& Astronomy
Baton Rouge, LA 70803-001, USA}

\author[0000-0002-6663-7675]{S. P. S. Eyres}
\affiliation{Faculty of Computing, Engineering \& Science, University of 
South Wales, Pontypridd, CF37 1DL, UK}

\author[0000-0003-1319-4089]{R. D. Gehrz}
\affiliation{Minnesota Institute for Astrophysics, University of Minnesota,\\
116 Church Street SE, Minneapolis, MN 55455, USA}

\author[0000-0003-1892-2751]{R. M. Wagner}
\affiliation{Department of Astronomy, The Ohio State University, 140 W. 18th Avenue, Columbus, OH 43210, USA}
\affiliation{Large Binocular Telescope Observatory, 933 North Cherry Avenue,
Tucson, AZ 85721, USA}

%%%% Optional Editorial Information
\correspondingauthor{C.E. Woodward}
\email{chickw024@gmail.com}
\received{2021 June 21}
%% \revised{2021 August 16}
\accepted{2021 August 17}
\published{To Appear in The Astronomical Journal}
%%\submitjournal{The Astronomical Journal}

%%%%
%012345678901234567890123456789012345678901234567890123456789012345678901234567890

\begin{abstract}

Luminous Red Variables (LRVs) are most likely eruptions that are the outcome of stellar mergers. 
V838~Mon is one of the best-studied members of this class, representing an archetype
for stellar mergers resulting from B-type stars. As result of the merger event, ``nova-like'' eruptions 
occur driving mass-loss from the system. As the gas cools considerable circumstellar dust is formed. 
V838~Mon erupted in 2002 and is undergoing very dynamic changes in its dust composition, geometry, 
and infrared luminosity providing a real-time laboratory to validate mineralogical condensation sequences 
in stellar mergers and evolutionary scenarios. We discuss recent NASA Stratospheric Observatory
for Infrared Astronomy (\sof{}) 5 to 38~\micron{} observations combined with 
archival NASA \spit{} spectra that document the temporal evolution of the freshly  formed (within the last $\ltsimeq 20$ yrs) 
circumstellar material in the environs of V838 Mon. Changes in the 10~\micron{} spectral region are strong evidence
that we are witnessing a  ‘classical’ dust condensation sequence expected to occur in oxygen-rich environments 
where alumina formation is followed by that of silicates at the temperature cools.

\end{abstract}

%% From astrothesaurus.org

\section{Introduction}\label{sec:intro}

%%%%
%012345678901234567890123456789012345678901234567890123456789012345678901234567890

Luminous Red Variables (LRVs) are characterized by very high luminosities, low effective temperatures, 
long ($\gtsimeq200$ day) evolution timescales of the eruption, and consequently large eruption 
energies \citep[see Fig.~1 in][]{2012PASA...29..482K}. They also display the presence 
of gas-phase AlO, SiO, SO, SO$_{2}$ and occasionally H$_{2}$S emission and/or absorption 
\citep{2018A&A...617A.129K}, and dusty circumstellar discs that show evidence of 
alumina (Al$_2$O$_3$) and other solid oxides \citep{2015ApJ...814..109B}. Their high luminosity at 
maximum \citep{2006AJ....131..984B} ($M_{\rm bol} \sim -10\, \rm{with}\, M_v \ltsimeq -9$, surpassing 
typical classical novae at maximum, $M_v \ltsimeq -8$) is confirmed by the detection  of LRVs in 
M31 \citep{2006AJ....131..984B} and in other galaxies \citep{2012PASA...29..482K, 2015ApJ...805L..18W, 2016MNRAS.458..950S}. 
While their nature was initially unclear (with nova eruptions, planet-swallowing stars, very late thermal pulses 
having been suggested) the ``best-bet'' scenario, based on V1309~Sco, is the merger of 
two stars \citep{2006A&A...451..223T, 2020ApJ...893..106M, 2021A&A...646A.119P}  within a triple or 
higher system \citep{2021arXiv210607427K}. \citet{2019A&A...630A..75P} present a recent 
extensive review of the phenomena.

%%%%
%012345678901234567890123456789012345678901234567890123456789012345678901234567890
V838~Mon is the best-studied LRV. Shortly after its eruption in 2002 \citep{2002IAUC.7785....1B}, 
a light echo -- due to reflection of the eruption off circumstellar dust -- became very prominent 
\citep{2003Natur.422..405B}; the echo was also prominent in \spit\ (+MIPS) \citep{2006ApJ...644L..57B} 
and \textit{Herschel}  \citep{2016A&A...596A..96E} imagery at $\lambda \gtsimeq 70~$\micron. 
V838~Mon is 6.2~kpc distant, a value tightly constrained from polarized light echo studies \citep{2008AJ....135..605S}.

Early in the eruption it displayed AlO, TiO and VO bands in the near-infrared 
\citep{2002A&A...395..161B, 2003MNRAS.343.1054E, 2004ApJ...607..460L}. 
\citet{2003MNRAS.343.1054E} classified V838~Mon as an ``L~supergiant,'' with an effective temperature 
T$_{\rm{eff}} \ltsimeq 2,300$~K.  Over the period 2002 through 2004 there were several absorption features 
present that are due to rotational-vibrational transitions in water \citep{2005ApJ...627L.141B}.
\citet{2005ApJ...627L.141B} concluded that the water arises from a cool $\sim 800$~K region, and that 
the excitation temperature and water column density were decreasing with time; this latter temperature 
was consistent with that deduced by \citet{2004ApJ...607..460L}. As of 2009 however, \citet{2015AJ....149...17L}
concluded that the ejected material was at a radial distance of $\simeq 263$~au and had temperature 
of 285~K. \citet{2014A&A...569L...3C} found that V838~Mon is surrounded by a flattened dusty 
structure (position angle $-10\degr$), that is likely transitory and extends to several hundred au from 
the central star based on mid-infrared interferometric imagery. \citet{2016A&A...596A..96E} found an 
extended source region of cold dust emission $\approx 2.7$~pc in size ($\sim$ 1\farcm5 in diameter) 
surrounding V838~Mon. Similar structures are seen post-AGB giant oxygen-rich systems, 
such as 89 Her \citep[][]{2014A&A...568A..12H}, using interferometric techniques.

%%%%
%012345678901234567890123456789012345678901234567890123456789012345678901234567890
Here we present recent observations of V838~Mon obtained with the NASA Stratospheric Observatory
for Infrared Astronomy \citep[SOFIA;][]{2009AdSpR..44..413G, 2012ApJ...749L..17Y}. The objective 
was to investigate the nature and dynamic evolution of the system's dust that formed in the material 
ejected by the stellar merger. 

We find significant changes in dust chemistry in the circumstellar environment. 
Comparison of the recent \sof{ } measurement of the spectral energy distribution to prior \spit{ } spectra obtained
almost a decade earlier suggests that in the 10~\micron{} region we are observing signatures of 
a `classical' dust condensation sequence that is expected to occur in oxygen-rich 
environments \citep{1990fmpn.coll..186T, 2013A&A...560A..75K} where alumina (Al$_{\rm{2}}$O$_{\rm{3}}$) 
forms initially in the hot, T $\sim 1,700$~K dust envelope \citep{2000A&AS..146..437S} followed by 
the formation of various silicates at cooler temperatures of T $\simeq 1,200$~K
\citep{1998Ap&SS.255..415T, 2010pdac.book...27G}.

%%%%
%012345678901234567890123456789012345678901234567890123456789012345678901234567890

%% 
%% Temporary placement Table 1 (version#3)
%\input{table1}

%%%%%%%%%%%%%%
%% aastexv6.1
%% The * option informs LATEX that a table (or figure) will span both columns i
%% in a two column style.
%% \begin{deluxetable*}
%%     ...
%% \end{deluxetable*}

%%\begin{longarotatetable}

%%\end{longarotatetable}}
%%\movetabledown=-2.0in
%%\begin{longrotatetable}
%%%%%%%%%%%%%%%%%%%%%%%%%%%%%%%%%%%%%%%%%%%%%%%%%%%%
\begin{deluxetable*}{@{\extracolsep{0pt}}lccrrcc}
%%%%%%%%%%%%%%%%%%%%%%%%%%%%%%%%%%%%%%%%%%%%%%%%%%%%%
\tablenum{1}
% Next command reduces the cap between table columns
\setlength{\tabcolsep}{2pt} % General space between cols (6pt standard)
%
%
%\tablewidth{0pt}
\tablecaption{FORCAST Observational Summary -- V838 Mon\tablenotemark{$\dagger$}\label{tab:sobstab}}
%\tabletypesize{\scriptsize}
\tablehead{
 &&\colhead{Grism} &\colhead{Single}&\colhead{Total}  \\
\colhead{Mean}&&\colhead{or}&\colhead{Frame}&\colhead{On Source} \\
\colhead{Observation} & \colhead{Instrument} & \colhead{Filter} & \colhead{Exposure} &\colhead{Integration}  &\\
\colhead{2019 UT Date} &\colhead{Configuration} &\colhead{$\lambda_{eff}$} &\colhead{Time} &\colhead{Time} &\colhead{CALERR\tablenotemark{a}} \\
\colhead{(mm-dd hr:min:s)} &  &\colhead{($\mu$m)} &\colhead{(sec)}  &\colhead{(sec)} 
 }

\startdata
(FO\_F628)\\
10-23T09:36:29.9 & Imaging Dual & \phn7.7 & 25.66 & 359.20 & \nodata \\
10-23T09:58:02.8 & Imaging Dual  & 11.2  & 27.07 & 324.88 & \nodata  \\
10-23T09:36:29.9 & Imaging Dual & 31.5 & 26.34 & 640.18 & \nodata     \\
10-23T08:50:20.2 & Grism SWC    & G063  & 19.92  & 637.47 & 0.0348  \\[2pt]
(FO\_F629)\\
10-24T10:33:13.7 & Imaging Dual & 19.7  & 26.15 & 313.78 & \nodata  \\
10-24T10:33:13.7 & Imaging Dual & 37.1 & 26.15 & 313.78 & \nodata    \\
10-24T08:07:41.8 & Grism LWC    & G111  & 34.37 & 1031.17 & 0.0743  \\
10-24T08:52:02.0 & Grism LWC    & G227  & 34.31 & 1029.40 & 0.0055  \\
10-24T09:36:18.3 & Grism LWC    & G329  & 32.28 & 2388.36 & 0.0136 \\[2pt]
(FO\_F630)\\
10-25T09:40:11.9 & Grism SWC     & G063  & 13.34  & 213.39 & 0.0348 \\
10-25T10:02:45.7 & Grism LWC     & G111   & 32.57 & 586.20  & 0.0743  \\
\enddata
\tablenotetext{\dagger}{Data files are available through the Infrared Processing and Analysis Center (IPAC) Infrared Science
Archives (IRSA) at \url{https://dcs.arc.nasa.gov} }
\tablenotetext{a}{Pipeline systematic photometric calibration error for the grating. }
\end{deluxetable*}

%%%% 
% Insert Table2 photomtery revised table format (version #4)
%\input{table2}
%% \label{tab:phot-tab}

%% Table 2: input for tables in manuscript
%% Using photometry in 7pix radius aperture
%% which represents the average PSF radius for all the filter FWHM
%%
%
% Imaging Photometry Table for SOFIA V838 Mon data
%
%\documentclass[12pt,preprint,citecolor=blue]{aastex6}
%\begin{document}

%%%%%%%%%%%%%%
%% aastexv6.1
%% The * option informs LATEX that a table (or figure) will span both columns i
%% in a two column style.
%% \begin{deluxetable*}
%%     ...
%% \end{deluxetable*}
%% If one wants the table to span multiple columns in two-column format need the *
%%%%%%%%%%%%%%%%%%%%%%%%%%%%%%%%%%%%%%%%%%%%%%%%%%%%
\begin{deluxetable}{@{\extracolsep{0pt}}lcc}
%%%%%%%%%%%%%%%%%%%%%%%%%%%%%%%%%%%%%%%%%%%%%%%%%%%%%
\tablenum{2}
% Next command reduces the cap between table columns
\setlength{\tabcolsep}{5pt} % General space between cols (6pt standard)
%
%
%\tablewidth{0pt}
\tablecaption{V838 Mon Photometry\tablenotemark{$\dagger$}\label{tab:phot-tab} }
\tablehead{
&\colhead{Flux} &\colhead{ }\\
\colhead{Filter} &\colhead{Density} &\colhead{Flux}\\
\colhead{($\mu$m)} &\colhead{(Jy)} &\colhead{($\times 10^{-12}$ W m$^{-2}$)}
}
\startdata
\phn7.7 & \phn5.912 $\pm$ 0.059 & 2.302 $\pm$ 0.023 \\
11.2  & 19.416 $\pm$ 0.074 & 5.197 $\pm$ 0.020 \\
19.7  & 38.454 $\pm$ 0.088  & 5.852 $\pm$ 0.013 \\
31.5 & 26.161 $\pm$ 0.077 & 2.490 $\pm$ 0.007 \\
37.1 & 17.527 $\pm$ 0.266 & 1.416 $\pm$ 0.021 \\
\enddata

\tablenotetext{\dagger}{Measured in a circular aperture with a diameter of 10\farcs75 centroided on the photocenter of V838 Mon
in each \sof{} FORCAST image at a given filter.}
%\tablenotetext{\star}{In $\lambda F_{\lambda}$ space.}
%\tablenotetext{b}{Average of 9 sec acquisition images.}
%\tablecomments{ }
\end{deluxetable}

%%%%

\section{Observations and Data Reduction}\label{sec:obsec}

Mid-infrared observations of V838~Mon were conducted in 2019 October on three consecutive 
flight originating from Palmdale, CA with the SOFIA airborne observatory using the Faint 
Object InfraRed CAmera \cite[FORCAST;][]{2018JAI.....740005H},  the dual-channel mid-infrared imager 
and grism spectrometer operating from 5 to 40~\micron, mounted at the Nasmyth focus of the 
2.5-m telescope. V838 Mon was imaged (platescale of 0\farcs768 per pixel) in the 
mid-infrared in three filters, F7.7 ($\Delta\lambda = 0.47$~\micron)
narrow band, F11.2 ($\Delta\lambda = 2.7$~\micron), 
and F31.5 ($\Delta\lambda = 5.7$~\micron),  and the 
Short Wavelength  Camera (SWC) grism (G063) on the first flight, while on the 
second flight imaging in the F197.7 ($\Delta\lambda = 5.5$~\micron)
and F37.1 ($\Delta\lambda = 3.3$~\micron) filters was 
performed in addition to Long Wavelength Camera (LWC) grism observations with three 
gratings (G111, G227, and G329). On the third night the grism G063 and G111 observations were 
repeated. 

For all spectroscopic observations the instrument was configured using a long-slit 
(4\farcs7 $\times$ 191\arcsec) which yields a spectral resolution $R = \lambda/\Delta\lambda \sim$ 
140--300. The position angle of the slit was arbitrary. V838~Mon was imaged with short 9 sec 
exposures in the SWC using the F111 filter to position the target in the slit. Both imaging and 
spectroscopic data were obtained using a 2-point chop/nod in the Nod-Match-Chop (C2N) mode 
with 45\arcsec\, chop and 90\arcsec\,nod amplitudes at angles of 30$^{\circ}$/210$^{\circ}$ in the 
equatorial reference frame. Flight altitudes were $\simeq 13,100$~m.

Table~\ref{tab:sobstab} summarizes the all observational data sets discussed herein.

%%%%
%012345678901234567890123456789012345678901234567890123456789012345678901234567890

\subsection{SOFIA Spectra}

\begin{figure*}[!ht]
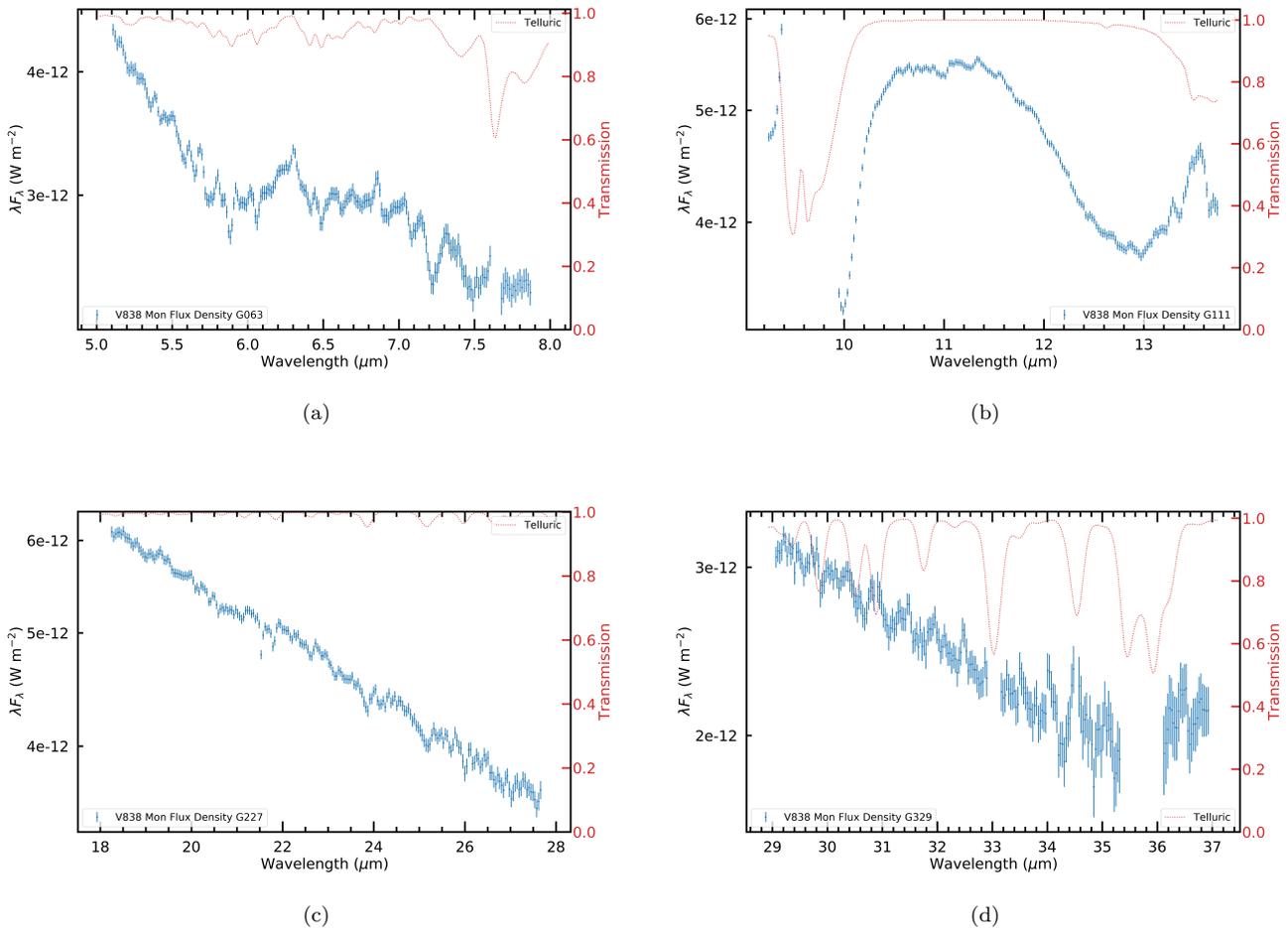

\figurenum{1}
\gridline{
\rotatefig{0}{fig1a.png}{1.0\columnwidth}{(a)}
\rotatefig{0}{fig1b.png}{1.0\columnwidth}{(b)}
}
\gridline{
\rotatefig{0}{fig1c.png}{1.0\columnwidth}{(c)}
\rotatefig{0}{fig1d.png}{1.0\columnwidth}{(d)}
         }
\caption{\label{fig:obs_each_grating_alone} V838~Mon SOFIA FORCAST spectra shown by individual 
grating to highlight spectral details and the signal-to-noise quality of the data. The panels are 
(a) G063, (b) G111, (c) G227, and (d) G329. The G063 and the G111 spectra are averages of two 
different observations on separate and distinct flight missions (the difference in the spectral
calibration were within the CALERR uncertainties (Table~\ref{tab:sobstab}). The uncertainties 
at a given spectral data point were propagated in quadrature. The dotted red line depicts the 
model atmospheric transmission at the flight altitude of the observations. Gaps in the contiguous 
spectral coverage arise from regions where the atmospheric transmission was modeled to 
be $\ltsimeq 70$\%.}
\end{figure*}

The FORCAST scientific data products were retrieved from the Infrared Processing and Analysis 
Center (IPAC) Infrared Science Archives (IRSA) after standard pipeline processing and flux calibration 
was performed \citep[for details see][]{2015ASPC..495..355C}. Computed atmospheric 
transmission models for the flight altitudes (which are contained in the data products) were used to 
mask-out grism data points in wavelength regions where the transmission was less than 
70\%. Spectra with grisms G063 and the G111 were obtained on two separate and distinct 
flight missions. However, the spectral energy distributions (SEDs) did not vary in 
shape or average intensity between the flights (i.e., the source was not detected to be varying on a 
timescale of $\ltsimeq 72$~hrs) and the difference in the spectral calibration were within the pipeline 
CALERR (systematic) uncertainties. Hence these data were averaged into a single spectrum for
each grating. Figure~\ref{fig:obs_each_grating_alone} presents panels for each individual 
grating segment, spanning their respective spectral free-range, to illustrate  details of the observed SED.

%\clearpage
%%
%% INSERT FIG1
%%  Insert individual Zoomed postage stamp spectra from originally created in the directory
%% Revised on : Mon Aug  9 12:11:34 CDT 2021
%% Figures created in python notebook FOAll_V3_V838Mon_ALLGRATINGS.IPYNB
%% Placed in RevRef1 subdirectory /figures as original copy
%% bash-3.2$ ls Fig1*2108*.png
%% Fig1a_V838Mon_210808_LogLin_Wm2_um_SOFIA_G063.png       Fig1c_V838Mon_210808_LogLin_Wm2_um_SOFIA_G227.png
%% Fig1b_V838Mon_210808_LogLin_Wm2_um_SOFIA_G111.png       Fig1d_V838Mon_210808_LogLin_Wm2_um_SOFIA_G329.png
%% bash-3.2$ 

%\vspace{-1.0cm}
%%
%1234568901234567890123456789012345678901234567890123456789012345678901234567890

%012345678901234567890123456789012345678901234567890123456789012345678901234567890
\subsection{SOFIA Imagery}

Images of V838~Mon were obtained on two different nights (see Table~\ref{tab:sobstab}).
Azimuthally averaged radial profiles of V838~Mon in each filter exhibited some evidence
of extended emission beyond the point-spread function (PSF) of 
point sources observed with FORCAST under optimal telescope jitter performance in each 
filter.\footnote{\url{http://www.sofia.usra.edu/Science/ObserversHandbook/FORCAST.html}} 
The mean FWHM of the azimuthally averaged radial profiles of the V838 Mon image data
was 3\farcs00 $\pm$ 0\farcs27 (i.e., 3.91 $\pm$ 0.35 pixels). Centroiding on the photocenter 
of V838~Mon, photometry in an effective circular aperture of diameter of 10\farcs75 
(i.e., a photometric aperture equivalent to $\simeq 3 \times$ the observed source
full width half maximum [FWHM]), with a background aperture annulus of inner 
radius 12 pixels (9\farcs22) and outer radius of 17 pixels (13\farcs01) was performed on the 
Level 3 pipeline co-added (*.COA) image data products using the Aperture Photometry Tool 
\citep[APT v2.4.7;][]{2012PASP..124..737L}.   Sky-annulus median subtraction 
\citep[ATP Model B as described in][]{2012PASP..124..737L} was used in the computation of the 
source intensity. The random source intensity uncertainty was computed using a depth of
coverage value equivalent to the number of co-added image frames. The calibration factors (and 
associated uncertainties) applied to the resultant aperture sums were included in the Level 3 
data distribution and were derived from the weighted average calibration observations of 
$\alpha$~Cet or $\alpha$~Tau. The resultant SOFIA photometry is presented in Table~\ref{tab:phot-tab}. 

\begin{figure*}[!ht]
\figurenum{2}
\begin{center}
\gridline{
\hspace{-1.5cm}
\rotatefig{0}{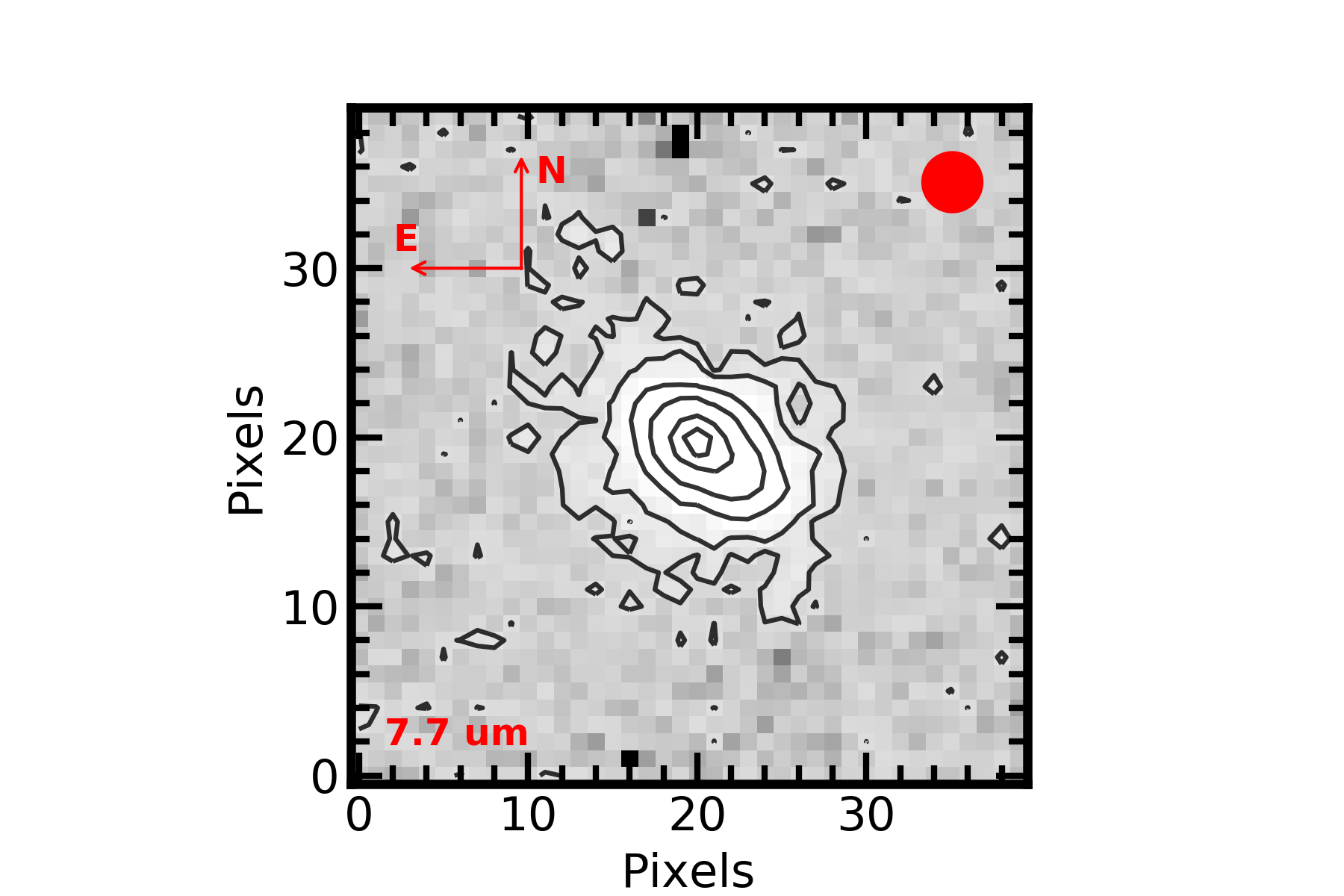}{0.45\textwidth}{(a)}
\hspace{-1.8cm}
\rotatefig{0}{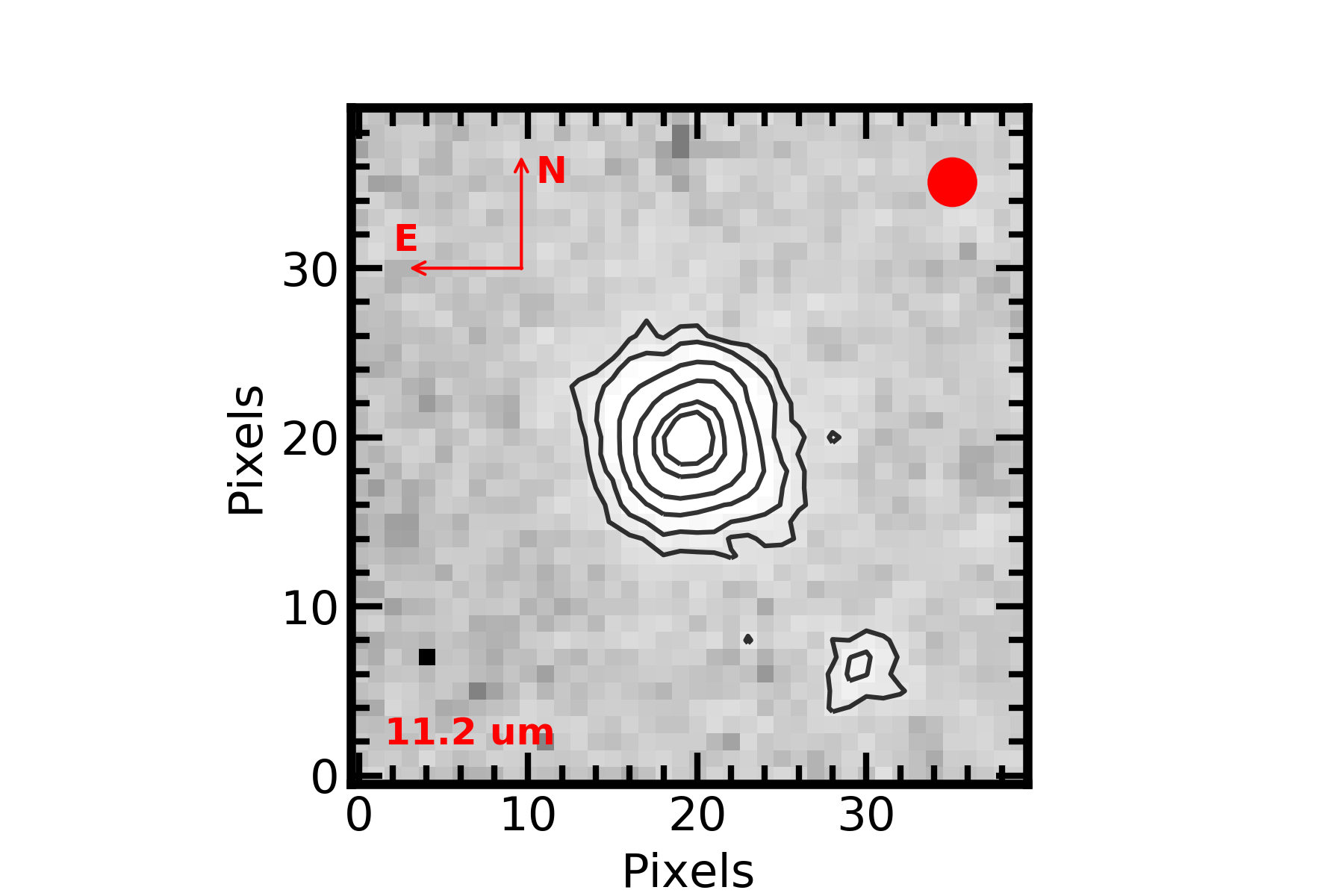}{0.45\textwidth}{(b)}
\hspace{-1.8cm}
\rotatefig{0}{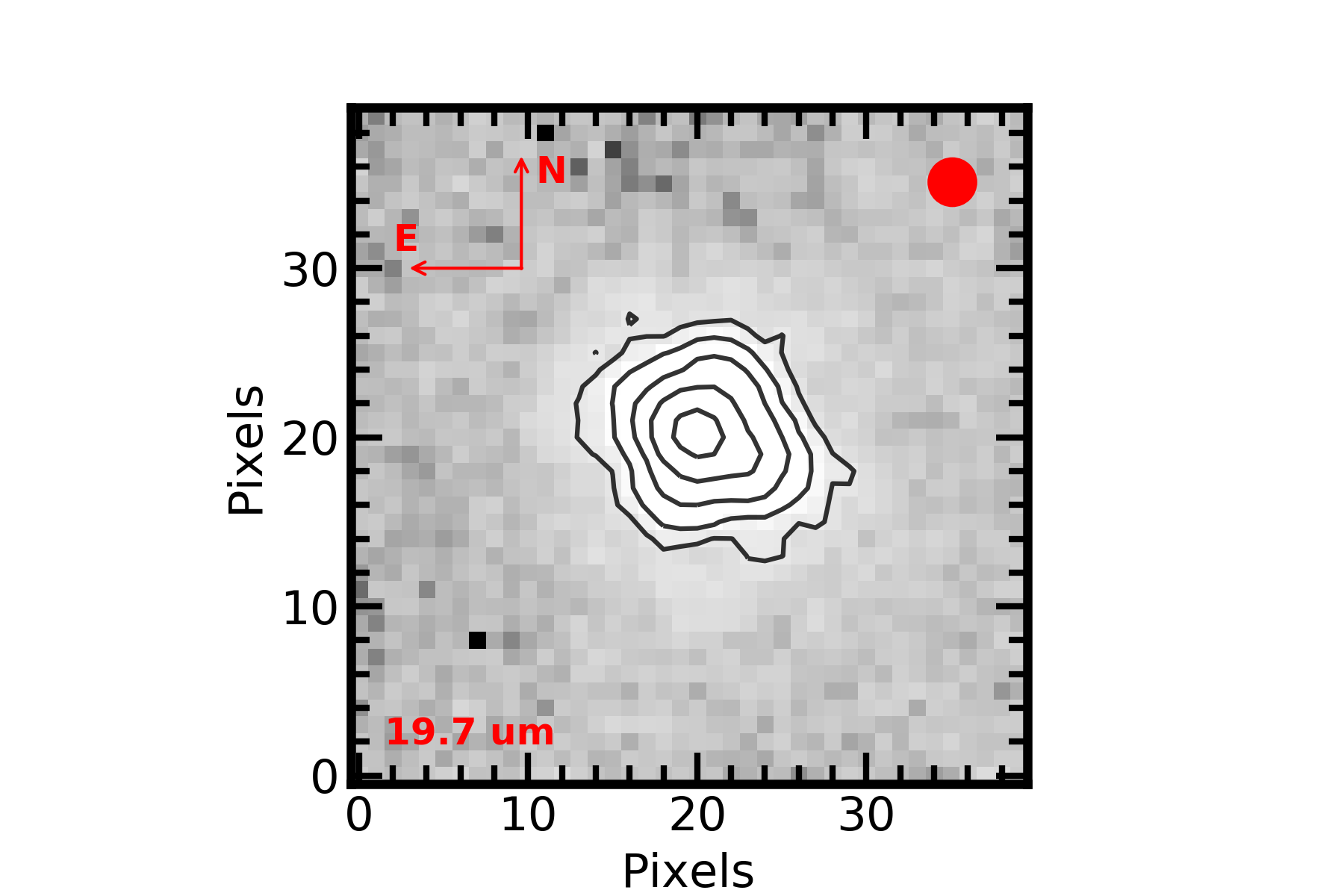}{0.45\textwidth}{(c)}
}
\gridline{
\hspace{1.0cm}
\rotatefig{0}{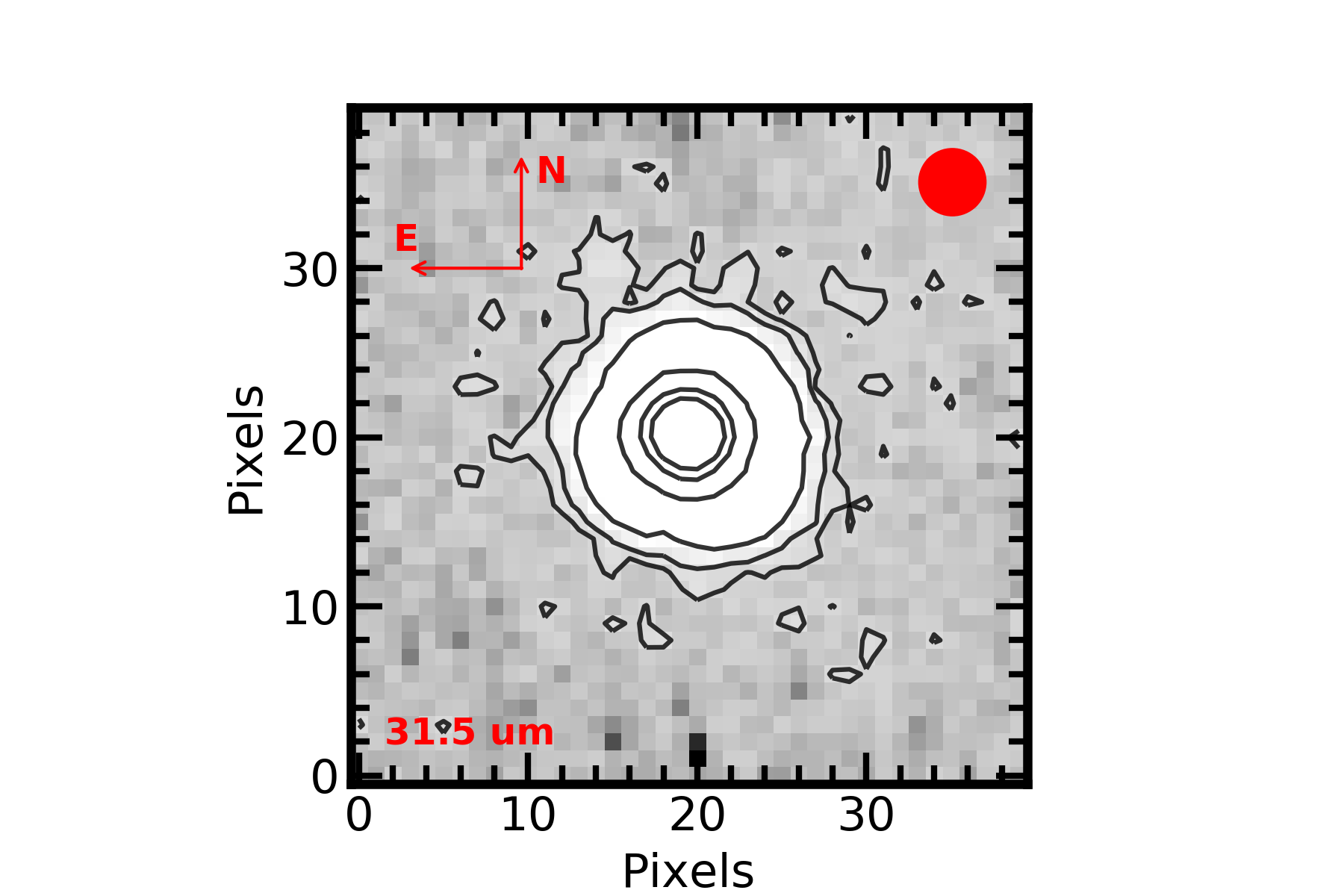}{0.45\textwidth}{(d)}
\hspace{-1.5cm}
\rotatefig{0}{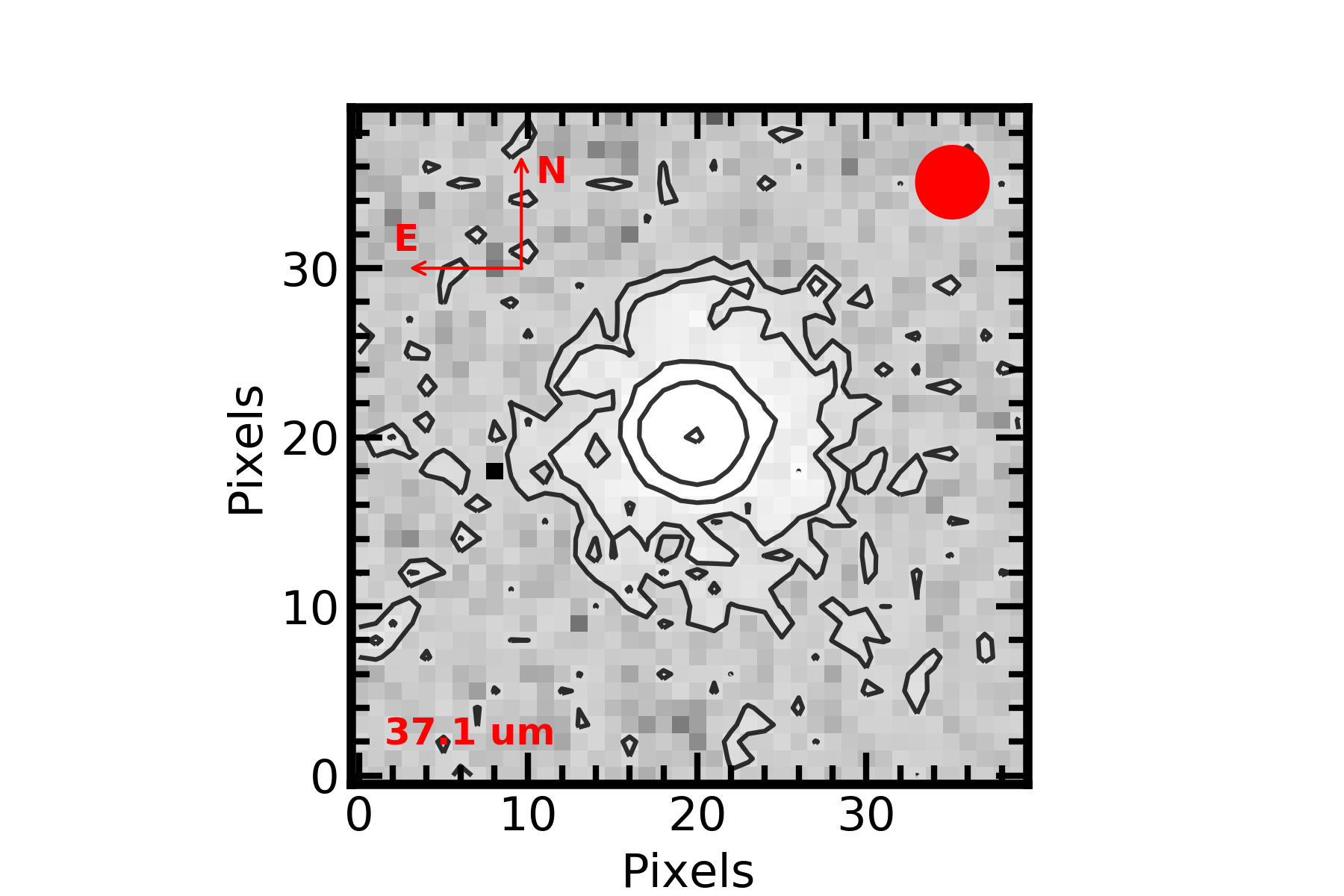}{0.45\textwidth}{(e)}
         }
\caption{\label{fig:obsflx_sofia_all_filters} SOFIA FORCAST gray-scale images of a 
30\farcs72 $\times$ 30\farcs72 (camera platescale 0\farcs768 per pixel) field-of-view centered on 
V838~Mon with superposed isophotal surface brightness contours, where the outermost contour 
is $4\sigma$ above the median sky background (Jy per pixel) measured off-source. With FORCAST, 
diffraction limited imaging is possible for $\lambda \geq 15$~\micron, limited however by telescope tracking 
and jitter. The FWHM PSF for each filter is given by the filled red circle in the upper right corner of each panel. 
(a) Filter F7.7, with contours at 0.0028 ($4\sigma$), 0.0062,  0.0173, 0.0518, 0.1725, and 0.2829 Jy. 
(b) Filter F112, with contours at 0.0060 ($4\sigma$),  0.0135, 0.0375, 0.1125, 0.3750, and 0.6160 Jy.
(c) Filter F197, with contours at 0.0200 ($4\sigma$), 0.0450, 0.1250, 0.3750, and 1.2500 Jy. (d) Filter F315
with contours at 0.0055 ($4\sigma$), 0.0124, 0.0345, 0.1034, 0.3450, and 0.5658 Jy. (e) Filter F371 with 
contours at 0.0112 ($4\sigma$), 0.0252, 0.0700, 0.2100, and 0.7000 Jy.}
\end{center}
\end{figure*}

Figure~\ref{fig:obsflx_sofia_all_filters} provides 30\farcs72 $\times$ 30\farcs72 postage-stamp 
gray-scale images with superimposed surface brightness contours. Generally the images
are point like, although the 7.7 through 19.7~\micron{} images are slightly elongated at
low surface brightness, with a position angle (PA; East of North) of $\sim 56$\degr. This elongation may be
associated with bipolar lobes of dust emission, perpendicular to the flattened-disk
(derived major axis size 23 mas at 8~\micron{} and 70 mas at 13~\micron)
structure interferometrically detected from 8 to 13~\micron{} by \citet{2014A&A...569L...3C}
that has a major axis PA of -10\degr.  Given the FORCAST
platescale and the beam FWHM (which has telescope jitter effects), \sof{} would not be able to
directly detect such a structure even at $\lambda \geq 30$~\micron. However, the \sof{} elongation
in the low surface brightness emission is similar in the position angle to that of the SiO maser 
emission channel velocity maps observed after 2018 November 20 \citep{2020A&A...638A..17O}. 
The 11.2~\micron{} \sof{} image also has a secondary source (likely a background source) 
12\farcs60 to the south-west of V838~Mon with a flux density (in a 10\farcs75 diameter aperture) 
of $1.23 \pm 0.07$ Jy.

%1234568901234567890123456789012345678901234567890123456789012345678901234567890
%%\clearpage
%%
%% INSERT FIG2
%% Postage stamp 30x30 arcsec images
%% Revised on : Mon Aug  9 12:11:34 CDT 2021
%% Figures created in python notebook ImageGraphics_for_SOFIA_Filters_v2.IPYNB
%% Placed in RevRef1 subdirectory /figures as original copy
%% bash-3.2$ ls Fig2*2108*.png
%% Fig2a_210809_image3030_f077+6contours.png       Fig2c_210809_image3030_f197+6contours.png       
%% Fig2b_210809_image3030_f112+6contours.png       Fig2d_210809_image3030_f315+6contours.png
%% Fig2e_210809_image3030_f371+6contours.png
%% bash-3.2$ 
%%
%% Fiddle around to get large enough panels spaced correctly in page over two-columns

%%%%%%%%%%%%%%%%%%
\subsection{\spit{} Spectra}

To study the long term spectral evolution of the circumstellar material, reduced archival \spit{} Infrared
Spectrograph \citep[IRS;][]{2004SPIE.5487...62H} high resolution
spectra (optimal difference extraction of nod1 and nod2) of V838~Mon also were retrieved from the 
Combined Atlas of Sources with Spitzer IRS Spectra\footnote{\url{https://cassis.sirtf.com/atlas/}}
\citep[Cassis,][]{2015ApJS..218...21L,2011ApJS..196....8L} from high spectral resolution
observations conducted on 2005 March 17.6139 (AORKey 10523136) and 
2008 December 10.0133 (AORKey 2543355).

\subsection{Ancillary optical and near infrared photometry}

Subsequent to outburst and initial decline of the light curve towards quiescence, optical BVIR 
photometry of V838~Mon acquired since JD 245 7648.91 (2016 Sept 17)
through JD 245 933.53 (2021 April 26) in the AAVSO\footnote{Observations from the AAVSO 
International Database, \url{https://www.aavso.orgdatabase}} database \citep{2020EPSC...14..314K}
show that there has been little ($\ltsimeq 0.5$~mag) change in the light curve; it has remained essentially
flat at all bands. 

V838 Mon was observed with The Nordic Optical Telescope's near-infrared camera and spectrograph (NOTCam), 
using its high resolution camera (0\farcs079 per pixel$^{-1}$) \citep{2000SPIE.4008..714A} and the broad-band 
filters $J$, $H$, and $K_{s}$ filters on 02 March 2020, JD 245 8911.43715 (the mid-point of the
observations), a few months after the 2019 SOFIA flights. Photometric calibration was performed using three 
2MASS stars in the field-of-view of the images and standard fields at a similar airmass 
observed just before the target. Standard infrared imaging reduction techniques using 
IRAF\footnote{IRAF is distributed by the National Optical Astronomy Observatories, which are 
operated by the Association of Universities for Research in Astronomy, Inc., under cooperative 
agreement with the National Science Foundation.} and apertures photometry (7\farcs0 circular diameter). 
The error (in magnitudes) is dominated by the uncertainty in the calibration stars. 

The observed near-infrared photometry, J = $6.59 \pm 0.05$, H $= 5.55 \pm 0.05$, K$_{s} = 4.76 \pm 0.05$, and
AAVSO photometry from JD 245 8879.565 of B $15.65 \pm 0.07$, V$=13.31 \pm 0.04$, R$= 11.47 \pm 0.03$, and
I$=9.47 \pm 0.01$ were de-reddened adopting a $E(B-V) = 0.87$ \citep[see][]{2015AJ....149...17L} with a 
standard galactic extinction curve \citep{1985ApJ...288..618R}. The de-reddened photometric data are 
used later in the analysis to constrain the SED at short wavelengths.

%012345678901234567890123456789012345678901234567890123456789012345678901234567890

\section{Discussion}\label{sec:disc}

%012345678901234567890123456789012345678901234567890123456789012345678901234567890
%%%%%%%%%%%%%%%%%%%%%%%%%%%%%%%%%%%%%%%%%%%%%%%%%%%%
The process of mass-loss and dust condensation is unknown for mergers, and the synoptic study (temporal
periods of several 100s of days to 10s of years) of V838~Mon may provide the constraints 
to confront observations with theoretical predictions. The nature and \textit{dynamic evolution} of the dust 
that forms in the material ejected by a stellar merger is not well-understood. Observations of V4332~Sgr 
(another proposed stellar merger) suggest that the grains have a alumina component
\citep{2007ApJ...666L..25B, 2015ApJ...814..109B}, which would be consistent with the strong AlO features
in oxygen-rich environments.

\subsection{The SOFIA 2019 Spectra}

The \sof{} spectra (Fig.~\ref{fig:obs_each_grating_alone}) exhibit interesting details regarding the SED 
of V838~Mon. At wavelengths longwards of 18~\micron{} the spectra are devoid of any strong 
line emission from hydrogen, helium, or forbidden lines such as [\ion{O}{4}] 25.91~\micron{} which is 
a strong coolant present in the late evolution of novae when electron densities in the ejecta are 
less than $\simeq 10^{6} - 10^{7}$~cm$^{-3}$ \citep{2015ApJ...812..132G, 2012BASI...40..213E, 2012ApJ...755...37H}. 
No molecular absorption bands or broad features from dust are evident. For example broad amorphous
silicate dust emission near 18~\micron{} or (Mg, Fe)O features near 19.5~\micron{} \citep{2002A&A...393L...7P}
are not present. 

%012345678901234567890123456789012345678901234567890123456789012345678901234567890
%%%%%%%%%%%%%%%%%%%%%%%%%%%%%%%%%%%%%%%%%%%%%%%%%%%%
The 8 to 14~\micron{} segment of the SED of V838~Mon is complex, being dominated by dust features, 
including a deep silicate absorption band centered near 10~\micron, an 
amorphous alumina (Al$_{2}$O$_{3}$) emission feature near $\sim 11.3$~\micron, and a
13~\micron{} feature that may be evidence of high temperature  spinel
\citep[MgAl$_{2}$O$_{4}$:][]{1999A&A...352..609P,  2013A&A...553A..81Z}. However, the measured 
FWHM of this latter feature ($\lambda_{o} =13.53$~\micron) is of order $\sim 0.1$~\micron, which is much 
narrower than the bandwidth measurements of high temperature ($300 \ltsimeq \rm{T(K)} \ltsimeq 928$) spinels 
\citep[Table 8 in][]{2013A&A...553A..81Z}. In addition, expected weaker 32~\micron{} spinel
emission bands are not evident in the \sof{} data.  Measurement of the 10~\micron{}
feature depth, $\tau_{9.7}$ is challenged by regions of poor atmospheric transmission in the \sof{} SED.
A more detailed discussion and modeling of the SED is discussed below (\S~\ref{sec:dustymods}).

\begin{figure}[!ht]
\figurenum{3}
\begin{center}
\includegraphics[trim=0.5cm 0.5cm 1.5cm 3.0cm,clip,width=1.0\columnwidth]{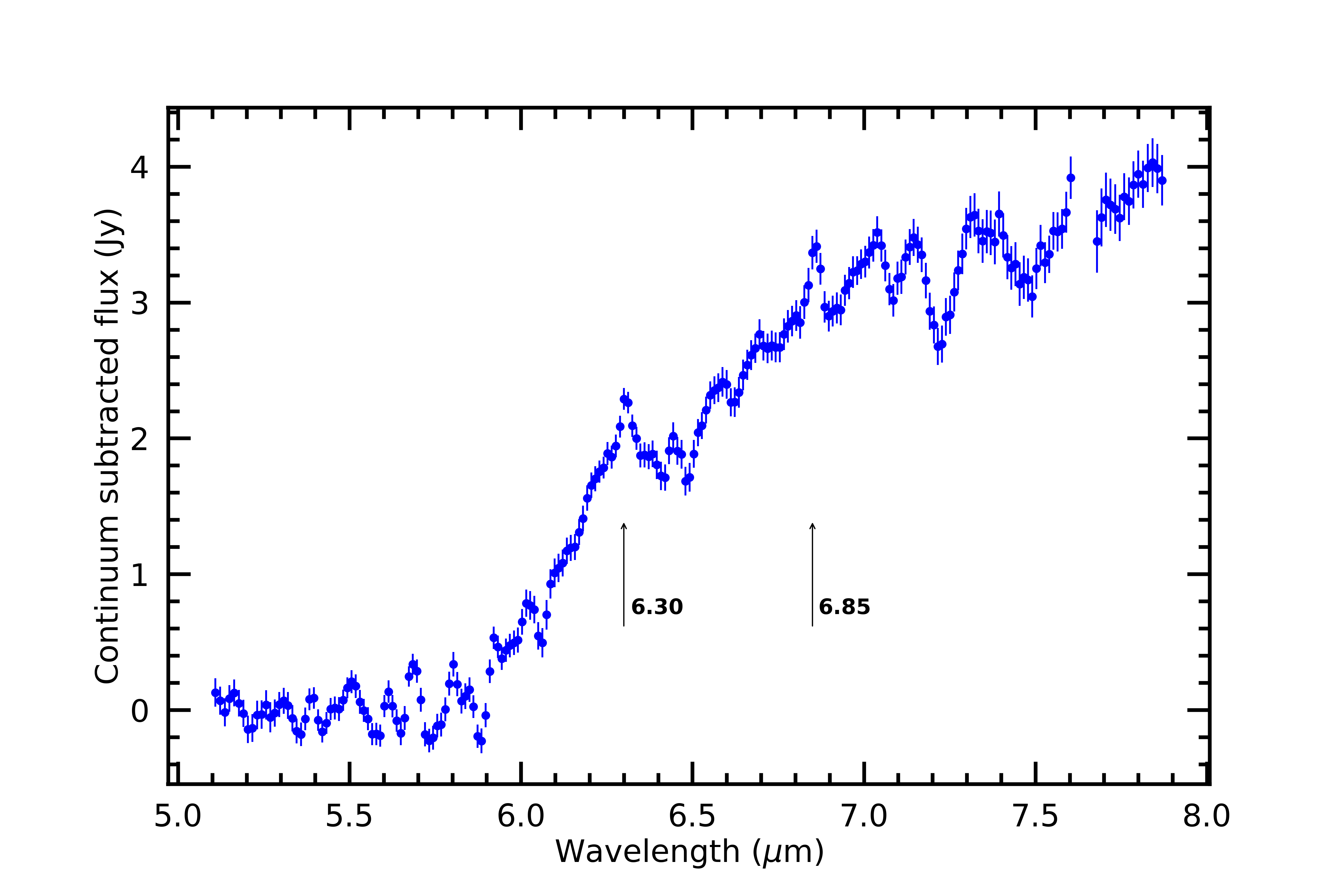}
\caption{\label{fig:obsflx_g063water} Continuum subtracted \sof{ } FORCAST spectra of V838 Mon from 5.0 to 8.0~\micron. 
A second-order Chebyshev polynomial fit using wavelengths from 5.2 to 5.6~\micron, that  are on the Rayleigh-Jeans 
tail of the stellar 2,300 K blackbody, was used to determine the local continuum. Features are indicated by the arrows.}
\end{center}
\end{figure}

From 5.0 to 8.0~\micron{} (Fig.~\ref{fig:obs_each_grating_alone}a), the SED is a composite of emission from
the Rayleigh-Jeans tail of the stellar 2,300~K blackbody and emission from a cooler dust component. Superposed
on the continuum there are a few suggestive emission features. Figure~\ref{fig:obsflx_g063water} shows the 
continuum subtracted residual emission, highlighting potential emission features. Features near 6.30~\micron{} and
6.85~\micron{} have been associated with water vapor emission ($\nu_{2}$ bands) and formaldehyde (H$_{2}$CO) 
in spectra of the dense circumstellar disk environments of T Tauri stars \citep{2014ApJ...792...83S}. Higher spectral
resolution observations with instruments like EXES \citep{2018JAI.....740013R} on SOFIA or 
MIRI on the James Webb Space Telescope (JWST) are necessary to confirm these identifications.

%1234568901234567890123456789012345678901234567890123456789012345678901234567890
%\clearpage
%%
%% INSERT FIG3
%% Revised on : Mon Aug  9 12:11:34 CDT 2021
%% Figures created in python notebook FOALL_V3_V838MON_ALLGRATINGS.IPYNB
%% Placed in RevRef1 subdirectory /figures as original copy
%%
%% bash-3.2$ ls Fig3*21*.png
%% Fig3_V883Mon_210808_LinLin_Jy_um_Annotate_2ndOrder_CntFitSub-SOFIA_G063.png
%% bash-3.2$ 
%%
% syntax: \includegraphics[trim=left bottom right top, clip]{file}
%%
%%%%%%%%%%%%%%%%%%%%%%%%%%%%%%%%%%%%%%%%%%

%1234568901234567890123456789012345678901234567890123456789012345678901234567890
\subsection{SED evolution and Dust Emission}

\begin{figure}[!htp]
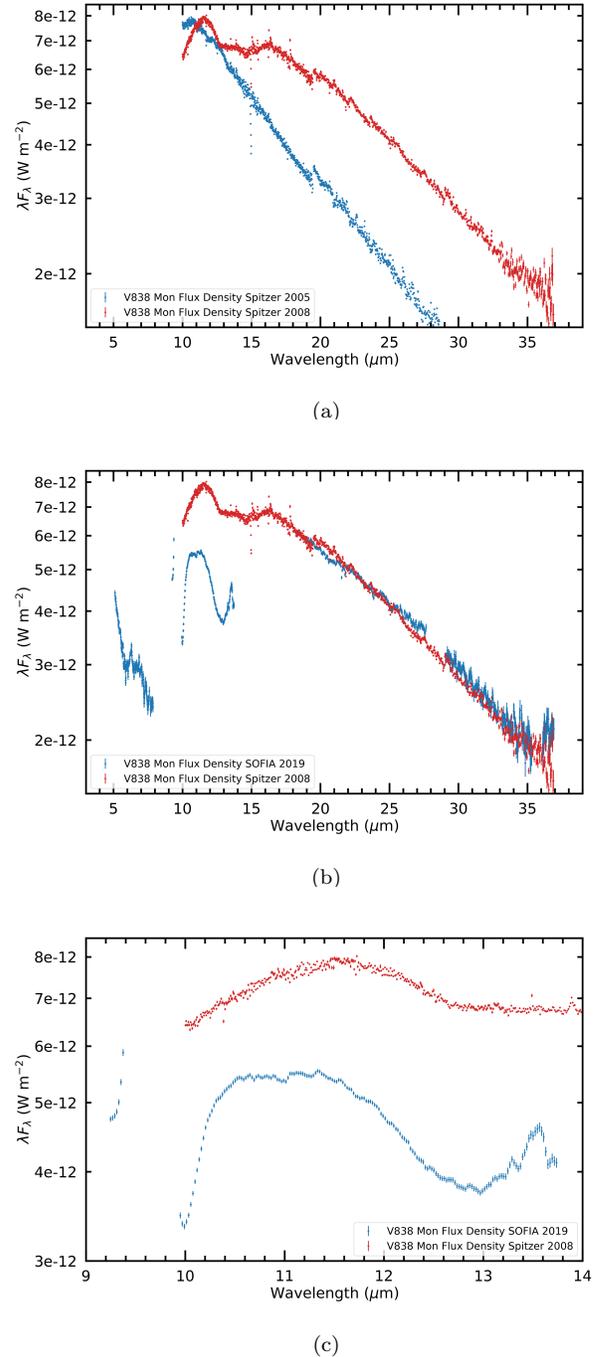

\figurenum{4}
\vspace{-0.4cm}
\gridline{
\rotatefig{0}{fig4a.png}{1.0\columnwidth}{(a)}
}
\vspace{-0.52cm}
\gridline{
\rotatefig{0}{fig4b.png}{1.0\columnwidth}{(b)}
}
\vspace{-0.52cm}
\gridline{
\rotatefig{0}{fig4c.png}{1.0\columnwidth}{(c)}
}
\caption{\label{fig:spitz05+08+sof19} Evolution of the infrared spectral energy distribution of 
V838~Mon. (a) The 2005 \spit{} IRS spectra (blue symbols) and the \spit{} 2008 IRS spectra 
(red symbols) show the slowly evolving spectral energy distribution, including the emergence of silicate 
emission bands, especially in the 10~\micron{} region between the two epochs. (b) Comparison of the 
composite 2019 \sof{} 5 to 36~\micron{} FORCAST spectra (blue symbols) and the 2008 \spit{} 10 
to 40~\micron{} IRS spectra (red symbols). Gaps in the \sof{} SED are due to non-contiguous spectral 
coverage of the FORCAST grisms. (c) Same as (b) but highlighting the the region between
9 and 14~\micron{} in detail which shows the marked change of structure in the 10~\micron{} feature.}
\end{figure}

\spit{} spectra between 2005 and and 2008 show that the mid-IR SEDs is evolving as shown in 
Figure~\ref{fig:spitz05+08+sof19}a. In 2005 the SED is smooth with little evidence for any broad dust emission
features; a blackbody fit to the SED yields a dust temperature, $T_{\rm{bb}} = 425~\pm~1.2$~K.
No 10~\micron{} feature is evident although it is difficult to draw a definite conclusion 
because the spectrum is saturated below 10~\micron. Three years later, the SED has markedly 
evolved, broad emission features are present, and at long wavelengths ($\lambda \geq 20$~\micron) 
the SED has a contribution from a cooler dust component.

The 2019 \sof{} SED is similar to that observed by \spit{} in 2008 only at wavelengths 
$\geq 18.0$~\micron{} as shown in Fig.~\ref{fig:spitz05+08+sof19}b. The flux density in the
9 to 14~\micron{} region, Figure~\ref{fig:spitz05+08+sof19}b observed in 2019 by \sof{} 
has decreased by $\sim 38$\% and exhibits a distinct broad ($\Delta\lambda \simeq 2.2$~\micron) 
emission band, a very distinct 10~\micron{} absorption feature as illustrated in
Figure~\ref{fig:spitz05+08+sof19}c. 

%\clearpage
%% INSERT FIG4
%% Revised on : Mon Aug  9 12:11:34 CDT 2021
%% Figures created in python notebook BBFlts_v4_V838MON.IPYNB
%% Placed in RevRef1 subdirectory /figures as original copy
%% bash-3.2$ ls Fig4*2108*.png
%% Fig4a_V838Mon_210809_LogLin_Wm2_um_Spitzer_2005_vs_2008.png
%% Fig4b_V838Mon_210809_LogLin_Wm2_um_SOFIA_all2019_vs_Spitzer2008.png
%% Fig4c_V838Mon_210809_LogLin_Wm2_um_10-15SOFIA_all2019_vs_Spitzer2008.png
%% bash-3.2$ 
%%
%%
%1234568901234567890123456789012345678901234567890123456789012345678901234567890

%012345678901234567890123456789012345678901234567890123456789012345678901234567890
%%%%%%%%%%%%%%%%%%%%%%%%%%%%%%%%%%%%%%%%%%%%%%%%%%%%
\vspace{0.5cm}
\subsection{DUSTY Models}\label{sec:dustymods}
To characterize the observed changes due to dust formation and evolution in V838~Mon, we 
modeled the system using the radiative transfer code DUSTY-DISK, which is similar to the original 
DUSTY code \citep{1995ApJ...445..415I}, but incorporates an additional disk component, appropriate for 
the case of V838 Mon. 

For the modeling two grain compositions were considered, silicate grains with
optical properties described by \citet{1984ApJ...285...89D} and amorphous, porous 
alumina  \citep{1997ApJ...476..199B}. These are bare grains, with no ice coatings. Typical
spectral indicators for ice-mantled grains are not seen in V383Mon. The water ice feature
at 3.05~\micron{} (due to and O$-$H stretch mode)  is not seen \cite[see Fig. 2 in][]{2004ApJ...607..460L}
nor is the 6.02~\micron{} H$-$O$-$H bending mode detected in archival \spit{} low-resolution IRS
spectra (which are unsaturated at $\lambda \ltsimeq 8$~\micron) shortly after outburst. 
Inspection of the SOFIA spectra near 6.02~\micron{} (Fig.~\ref{fig:obsflx_g063water}) also shows
no signature of a broad ice absorption feature.

Grids of simple models which varied the relative ratios of these two grain
components were constructed, adopting a Mathis-Rumpl-Nordsieck \citep[MRN,][]{1977ApJ...217..425M}
grain size distribution, $N(a) \propto a^{-q}$ with $q = 3.5$ and a grain-size range  of 
$5.0 \times 10^{-3} \leq \rm{a_{grain} (\micron)} \leq 2.5 \times 10^{-1}$. The 
input radiation field was represented by a single 2,300~K Planckian (blackbody) source commensurate
with a L3 supergiant, having an effective temperature of 2,300~K \citep{2015AJ....149...17L}. A spherical
shell of dust was illuminated by this source, where the dust temperature at the shell inner boundary 
was set at 400~K having a dust density distribution described by a 
inverse power-law ($\alpha = 2$, assuming a constant wind scenario) with the shell 
extending 2.5 times the inner radius ($Y=2.5$). Added to this was a disk illuminated the same source with 
the temperature at the outer disk+envelope boundary set to 25~K, with no accretion. The grids also comprised
a range of optical depths, specified at 0.55~\micron{}, varying in step size from 0.01 to 0.1 
spanning $0.5 \leq \tau_{0.55} \leq 5.0$. A bolometric flux (scaling factor)
of $3.1 \times 10^{-11}$~W~cm$^{-2}$ was adopted \citep[see Appendix discussion in ][]{2012A&A...544A..35J}. 

%012345678901234567890123456789012345678901234567890123456789012345678901234567890
%%%%%%%%%%%%%%%%%%%%%%%%%%%%%%%%%%%%%%%%%%%%%%%%%%%%
\subsection{Model Outcomes and Interpretation}

Initial analysis of the mid-infrared observational data of V838~Mon 
suggest that the SED of the  system at present can be explained as the sum of at least two components.
The first is a cool central star at  $\sim 2300$~K which is likely the central remnant of the merger. 
The SED of this component behaves as a blackbody at  2,300~K with a dusty envelope 
of modest optical depth such that the emergent radiation, modeled under assumptions of spherical geometry plus
a disk, reasonably (in the $\chi^{2}$ sense) reproduces the SEDs. The second component is
emission from dust in the disk+envelope.

The emergence of a 10~\micron{} dust feature was first observed in the 2008 \spit{} observations.
Clearly, this silicate emission feature at $\sim 10$~\micron{} has arisen newly formed in the 
intervening 3 years since 2005 (Figure~\ref{fig:spitz05+08+sof19}). This 10~\micron{} feature could arise 
from a combination of silicate dust (peaking at 9.7~\micron) and alumina dust (peaking at 11.3~\micron). 
More sophisticated RT modeling is required to robustly conclude whether the feature is composed 
purely of silicates or alumina or a combination of both, and to determine the constraints on the grain size distribution power-law. 

%012345678901234567890123456789012345678901234567890123456789012345678901234567890
%%%%%%%%%%%%%%%%%%%%%%%%%%%%%%%%%%%%%%%%%%%%%%%%%%%%

\begin{figure}[!ht]
\figurenum{5}
\begin{center}
\vspace{-0.4cm}
\gridline{
\rotatefig{0}{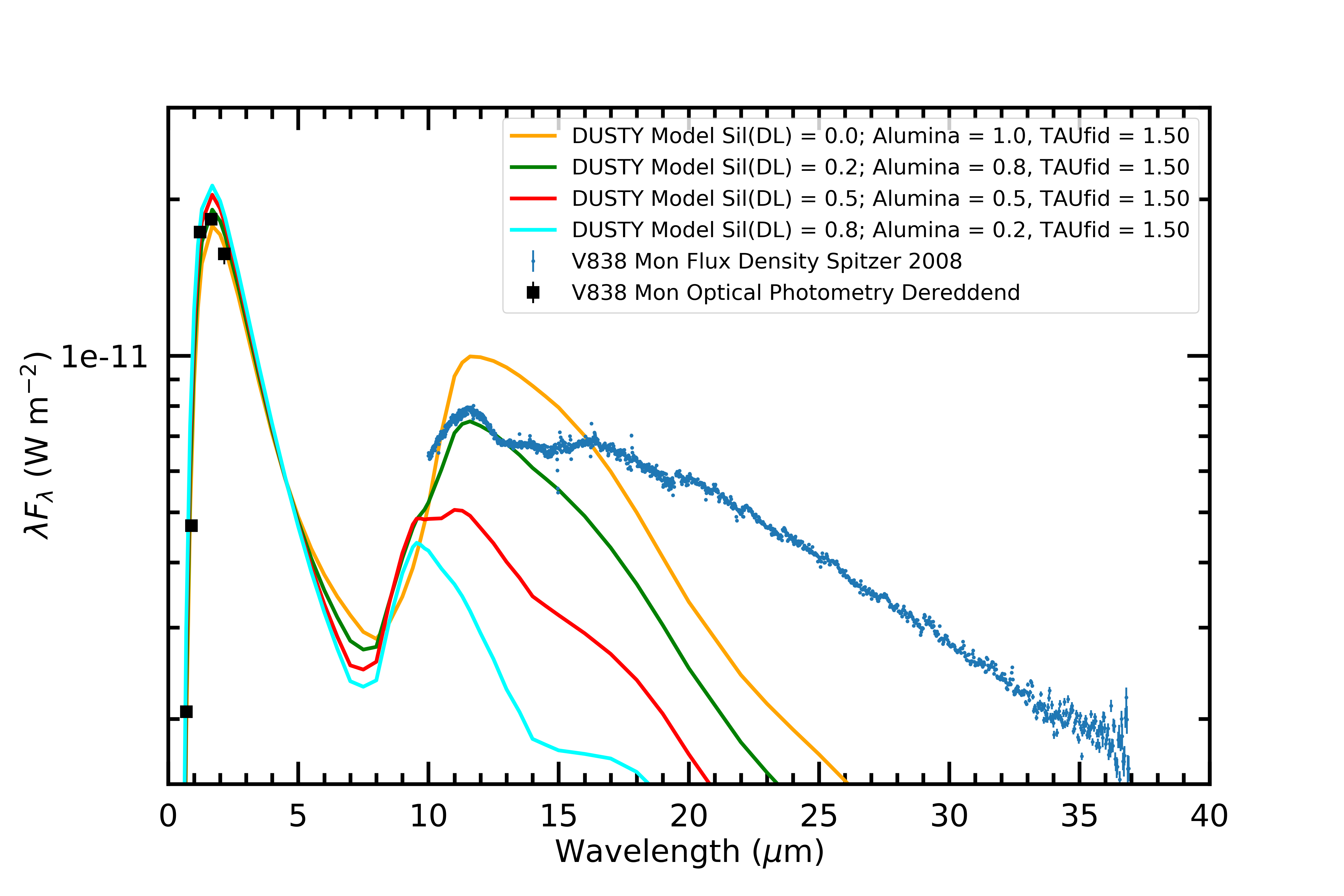}{1.0\columnwidth}{(a)}
}
\vspace{-0.52cm}
\gridline{
\rotatefig{0}{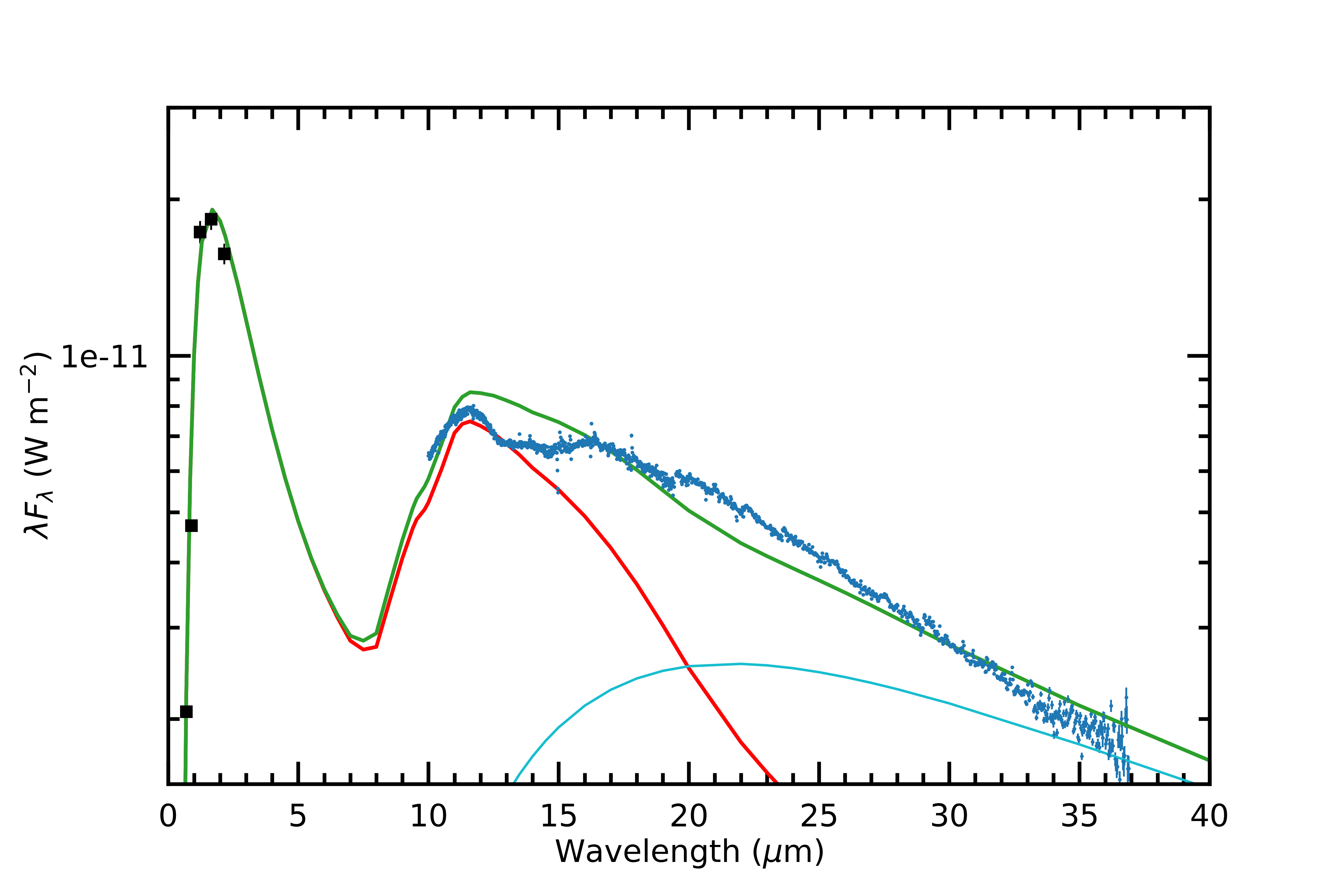}{1.0\columnwidth}{(b)}
}
\caption{\label{fig:sa_mixes_spit2008_dustymodels} DUSTY-DISK models of V838 Mon 2008 \spit{ } 
spectrum. The \spit{ } spectrum is depicted by the blue dots, while the de-reddend optical and infrared photometry 
are the black squares. (a) Representative sample of grid models illustrating the effect on varying the 
silicate-to-alumna dust ratios at a fixed optical depth $\tau_{0.55} = 1.50$. (b) Best-fit model composite
spectra (solid green line) that includes a grain mixture ratio Si:Alumina = 0.2:0.8,
emission from 2,300~K blackbody representing the stellar emission (red solid curve), and a third 
contribution to the composite SED from a $\simeq 170$~K blackbody (cyan line). The latter is necessary to 
account for the observed continuum emission at wavelengths $\gtsimeq 20$~\micron. The de-reddended
optical photometry is given by the black squares.}
\end{center}
\end{figure}

Figure~\ref{fig:sa_mixes_spit2008_dustymodels}a shows the 
DUSTY-DISK models with range of silicate-to-alumina ratio mixes (Si:Alumina) with $\tau_{0.55} = 1.50$ which 
illustrate how variation in the dust grain components alter the the shape and structure of the model SED. The 
best-fit to the shape of the 10~\micron{} region at this epoch is one with Si:Alumina = 0.2:0.8, suggesting that alumina 
dust dominates at this epoch (2008). In order to account for additional emission longward of 20~\micron{} a third 
component contributing to the overall model emission was necessary. This component is characterized
by thermal continuum emission likely from dust with a T$_{\rm{bb}} \simeq 170$~K
as shown in Figure~\ref{fig:sa_mixes_spit2008_dustymodels}b. This cooler component may be associated
with the cool circumstellar material detected by ALMA \citep{2021arXiv210607427K}. The sum of these 
three components gives a reasonable overall fit to the SED. However, the plateau between 
13 to 15~\micron{} in the observed \spit{} spectra could not be adequately reproduced by any 
combination of grain composition or size distributions.

%012345678901234567890123456789012345678901234567890123456789012345678901234567890
%%
%% INSERT FIG5
%% Revised on : Mon Aug  9 12:11:34 CDT 2021
%% Figures created in python notebook Busty_Output_parser_v3.IPYNB
%% Placed in RevRef1 subdirectory /figures as original copy
%%
%% bash-3.2$ ls Fig5*2108*.png
%% Fig5a_210808_man_sa_mixes+spitzer2008_dustymodel.png    Fig5b_210808_man_sa_mixes+bb+spitzer2008_dustymodel.png
%% bash-3.2
%% Insert models DUST_DISK Spitzer 2008
%%
%%%%%%%%%%%%%%%%%%%%%%%%%%%%%%%%%%%%%%%%%%%%%%%%%%%%%%
%012345678901234567890123456789012345678901234567890123456789012345678901234567890

The evolution of the 10~\micron{} region of the SED of V838~Mon has continued and demonstrates that the
chemistry of the circumstellar environment has changed over approximately the last decade. The 2019 \sof{} spectrum
shows that the emission plateau from 13 to 15~\micron{} has developed into a deep trough, the width of the
broad 10~\micron{} feature has narrowed ($\Delta\lambda \ltsimeq 2.0$~\micron) becoming more distinct, 
while an apparent absorption feature shortward of 10~\micron{} is seen (Figure~\ref{fig:spitz05+08+sof19}c).
This absorption may be a signature of silicates or more likely an artifact  caused by imperfect removal of the 
deep telluric feature that lies between 8 to 10~\micron{} (Figure~\ref{fig:obs_each_grating_alone}b).
In other oxygen rich environments, a deep 9.7~\micron{} absorption feature
is attributed to SiO materials and the depth of the feature indicates that silicates may now be the
dominant grain component. 

In other merger-system nova-likes that have dusty circumstellar envelopes, such as V1309~Sco, 
the broad spectral feature at 9.7~\micron{} is attributed to silicate grain solid-state 
absorption \citep{2013MNRAS.431L..33N}.  Our models of V838 Mon require amorphous silicates and the 
observed SED suggests a dust absorption feature indicative of an optically thick circumstellar environment
is present. Following the arguments discussed in \citet{2013MNRAS.431L..33N} an upper limit to the 
column density in V838 Mon can be derived from the observed depth of the 9.7~\micron{} feature 
(upper limit of $\sim 2.4 \times 10^{-12}$ W~$m^{-2}$) and the best-fit model continuum 
($\sim 4.6 \times 10^{-12}$ W~$m^{-2}$) at the same wavelength, which yields an optical depth 
$\tau_{9.7} \sim 0.3$. Using values of $Q_{\rm ext}$ and $Q_{\rm scat}$ for 'astronomical silicate' taken 
from \citet{1985ApJS...57..587D} and assuming amorphous silicate grains with radii $a$ between 0.1 to
3.0~\micron{} \citep[the upper limit to $a$ is set by the transition to a regime were the
9.7~\micron{} feature is suppressed, see Fig.~5 in][]{1993ApJ...402..441L} leads to a 
derived column density of between $\sim 8 \times 10^{8}$~cm$^{-2}$ to $2 \times 10^{10}$~cm$^{-2}$. 

\begin{figure}[!ht]
\figurenum{6}
\begin{center}
\includegraphics[trim=0.5cm 0.5cm 1.5cm 3.0cm,clip,width=1.0\columnwidth]{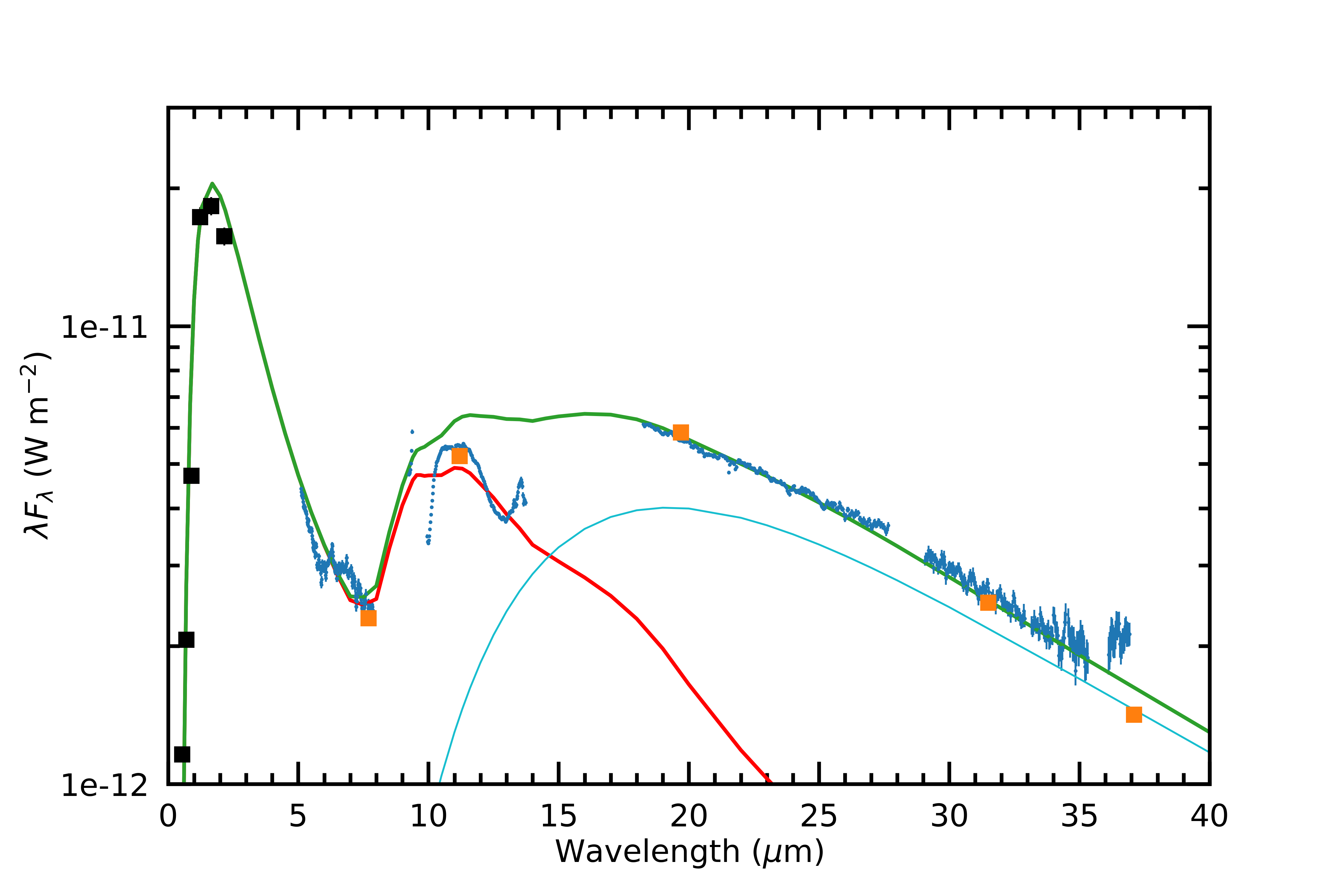}
\caption{\label{fig:sa_mixes_sof2019_dustymodels} DUSTY-DISK models of V838 Mon \sof{ } 
2019 spectrum. The \sof{ } spectra are depicted by the blue dots, while the de-reddend optical and infrared photometry 
are the black squares. The \sof{} photometry (Table~\ref{tab:phot-tab}) is indicated by the 
orange squares. Reasonable fit to the observed spectrum is achieved
with a silicate-to-alumina ratio of order 50:50, optical depth $\tau_{0.55} = 1.44$. The model composite
spectra (solid green line) includes model emission from the grain mixture, a 2,300~K blackbody representing 
the stellar emission (red solid curve), and requires a third contribution to the composite SED from a $\simeq 125$~K 
emissivity modified ($\epsilon_{\lambda} \propto \lambda^{-2}$) blackbody (cyan line). The latter contribution is necessary to 
account for the observed continuum emission at wavelengths $\gtsimeq 20$ \micron.}
\end{center}
\end{figure}

The rise of silicates also is supported by the shape and strength of
the broad 10~\micron{} band emission. This speculation is confirmed by DUSTY-DISK modeling of the 2019 \sof{}
composite spectra as shown in Figure~\ref{fig:sa_mixes_sof2019_dustymodels}. Models which best reproduce
the 10~\micron{} feature are those where the Si:Alumina ratio is now at least 50:50, with a slight decrease in the
optical depth to $\tau_{0.55} = 1.44$. A cooler third component with T$_{\rm{bb}} = 125$~K and a wavelength
dependent emissivity $\epsilon \propto \lambda^{-2}$ at wavelengths $\gtsimeq 10$~\micron{} are thought
to be present. The observed spectral evolution indicates that processing of the dust in V838 Mon is occurring,
perhaps similar to that in the environs of V1309 Sco \citep{2013MNRAS.431L..33N}. 
 
%012345678901234567890123456789012345678901234567890123456789012345678901234567890
%%
%% INSERT FIG6
%% Revised on : Mon Aug  9 12:11:34 CDT 2021
%% Figures created in python notebook Busty_Output_parser_v3.IPYNB
%% Placed in RevRef1 subdirectory /figures as original copy
%%
%% bash-3.2$ ls Fig6*2108*.png
%% Fig6_210808_man_sa_mixes+allphot+eps2bb+tau144+sc6.75e-12+sofia2019_dustymodel.png
%% bash-3.2$ 
%% Insert models DUST_DISK SOFIA 2019
%%
%%%%%%%%%%%%%%%%%%%%%%%%%%%%%%%%%%%%%%%%%%%%%%%%%%%%%%
%012345678901234567890123456789012345678901234567890123456789012345678901234567890

%012345678901234567890123456789012345678901234567890123456789012345678901234567890
%%%%%%%%%%%%%%%%%%%%%%%%%%%%%%%%%%%%%%%%%%%%%%%%%%%%
Clearly, there is a temporal evolution of this 10~\micron{}  feature with the relative strengths of the silicate 
and alumina components evolving with time in a manner consistent with the chaotic silicate hypothesis of 
mineral condensation described by \citet{1990Ap&SS.163...79N} or other models based on 
thermodynamically controlled evolution \citep{2000A&AS..146..437S}. The model fits 
(Figures~\ref{fig:sa_mixes_spit2008_dustymodels}, \ref{fig:sa_mixes_sof2019_dustymodels}) 
while not totally satisfactory, do permit two possible interpretations. First, there was a significant amount of 
alumina in the 10~\micron{} feature when this feature first developed. Modeling suggests that the Si:Alumina 
ratio was $\gtsimeq 0.5$. This would be observational evidence which supports the prediction that alumina 
should be the first dust condensate in an O-rich environment. Evidence for this prediction is rarely offered 
because most objects studied (e.g., Miras, AGB stars, etc.) are millions of years old. In the present case one is 
seeing this event happen in freshly condensed dust and almost in real-time. The temporal 
sequence of dust evolution in V383Mon may be a rare validation of mineralogical condensation 
sequences \citep{ 1990Ap&SS.163...79N} occurring in O-rich environments, perhaps only seen before in 
V4332~Sgr \citep{2007ApJ...666L..25B}.

Alternatively, one could conclude that the Sil:Alumina ratio changed between 2008 and 2019. It appears 
that the silicate fraction within the dust population has increased by $\gtsimeq 30$\%. The increase of silicate 
with time, at the expense of alumina can be explained as follows 
\citep[e.g.,][]{1990ApJ...350L..45S, 1990Ap&SS.163...79N}: in the initial stages, the higher reduction of Al with 
respect to Si leads to the preferential formation of Al-O bonds at the expense of Si-O bonds. This implies that 
the infrared bands of alumina associated with the Al-O stretching mode should be prominent early
in the formation of the chaotic silicates. However, as the Al atoms become fully oxidized, the higher abundance 
of Si will make the 9.7~\micron{} band associated with Si-O bonds dominate. 

The presence of alumina as a dust component  is not surprising.  V838 Mon is known to 
exhibit strong photospheric B-X infrared bands of AlO in the 
infrared \citep{2002A&A...395..161B, 2003MNRAS.343.1054E, 2004ApJ...607..460L}
which are present even today (as seen in the recent \sof{} spectra discussed herein).  Aluminum oxide 
is likely to play a significant role in the route to Al$_{2}$O$_{3}$ formation. LTE calculations 
by \citet{1999A&A...347..594G} show that any possible nucleation species that can go on to 
form dust around stars should begin with a monomer with exceptionally high bond energy. The 
AlO monomer satisfies this criterion and is thus a favored candidate to lead to the formation of 
larger Al$_{\rm{m}}$O$_{\rm{n}}$ clusters that serve as nucleation sites 
for the formation of other grains or to alumina grains themselves by homogeneous nucleation. 

We have not considered in our model the effect of ongoing processes such as  
annealing of the dust or grain growth. Annealing of silicate grains can  change the optical 
constants of the grain significantly,  as shown by the study of \citet{2000ApJ...535..247H}
and this can result in changes in the shape and peak of the silicate profile.  This point becomes 
relevant when comparing the evolution of dust features  across different epochs. The physics 
of grain growth in the expanding ejecta of novae, where the radiation field may be similar to the early
conditions in V838~Mon, is explored by \citet{2004A&A...417..695S}.

%012345678901234567890123456789012345678901234567890123456789012345678901234567890
\clearpage
\section{Summary}
Over the last decade the dust chemistry in the circumstellar environment of V838~Mon has dynamically evolved.
The temporal changes observed in the the 10~\micron{} is evidence of  a `classical'  dust condensation sequence
expected to occur in dense oxygen-rich regions. Further synoptic study of V838~Mon in the infrared with \sof{}
and JWST are required to explore timescales for condensation pathways, to ascertain the nature of the colder
component contributing to the spectral energy distribution at wavelengths $\gtsimeq 20$~\micron, and 
to understand the spatial distribution of the circumstellar emission.

% Note have the correct contract for SOFIA in extended mission
% See https://www.sofia.usra.edu/science/publications/information-authors

%012345678901234567890123456789012345678901234567890123456789012345678901234567890
%%%%%%%%%%%%%%%%%%%%%%%%%%%%%%%%%%%%%%%%%%%%%%%%%%%%
\acknowledgments
The authors would like to thank Zeljko Ivezi\'c for the helpful insight into the DUSTY and DUST-DISK
codes, and Floretin Millour for discussion of V838~Mon spectra in the 10~\micron{} region. 
The authors also appreciate the thoughtful critique and suggestions by the referee, 
which improved the manuscript. This work 
is based in part on observations made with the NASA/DLR Stratospheric Observatory for 
Infrared Astronomy (SOFIA). SOFIA is jointly operated by the Universities Space Research 
Association, Inc. (USRA), under NASA contract NNA17BF53C, and the Deutsches SOFIA 
Institut (DSI) under DLR contract 50 OK 0901 to the University of Stuttgart. Also 
based on observations made with the Nordic Optical Telescope, owned in collaboration by the 
University of Turku and Aarhus University, and operated jointly by Aarhus University, the University of Turku 
and the University of Oslo, representing Denmark, Finland and Norway, the University of Iceland and Stockholm 
University at the Observatorio del Roque de los Muchachos, La Palma, Spain, of the Instituto de Astrofisica
de Canarias. Financial support for this work and CEW was provided by NASA through award SOF07-0027 issued by USRA
to the University of Minnesota. DPKB is supported by a CSIR Emeritus Scientist 
grant-in-aid which  is being hosted by the Physical Research Laboratory, Ahmedabad. TL acknowledges 
financial support from the Czech Science Foundation (GA\,\v{C}R 20-00150S). We acknowledge with 
thanks the variable star observations from the AAVSO International Database contributed by observers 
worldwide and used in this research. RDG was supported by the United States Air Force. 

\facilities{NASA SOFIA (FORCAST), NASA \spit{} (IRS), AAVSO, NOT}

\software{IRAF, REDUX \citep{2015ASPC..495..355C}, APT \citep{2012PASP..124..737L}, astropy \citep{2018AJ....156..123A}}

%%%%%%%%%%%%%%%%%%%
%\vspace{8cm}
%\vskip=1.0cm
\clearpage
\bibliography{mainArXivV838mon}{}
\bibliographystyle{aasjournal}

\end{document}